\documentclass[a4paper,
              ]{jacow}
\makeatletter%
	\ifboolexpr{bool{xetex}}
	 {\renewcommand{\Gin@extensions}{.pdf,%
	                    .png,.jpg,.bmp,.pict,.tif,.psd,.mac,.sga,.tga,.gif,%
	                    .eps,.ps,%
	                    }}{}
\makeatother

\ifboolexpr{bool{xetex} or bool{luatex}} 
 {}                                      
 {\usepackage[utf8]{inputenc}}           

\usepackage[USenglish]{babel}			 

\usepackage[final]{pdfpages}
\usepackage{multirow}
\usepackage{ragged2e}
\usepackage{lineno}
\ifboolexpr{bool{jacowbiblatex}}%
 {%
  \addbibresource{jacow-test.bib}
  \addbibresource{biblatex-examples.bib}
 }{}
\begin{document}
\title{UPGRADE OF THE TWO-SCREEN MEASUREMENT SETUP IN THE AWAKE EXPERIMENT}

\author{M. Turner\thanks{marlene.turner@cern.ch}\textsuperscript{1}, V. Clerc, I. Gorgisyan, E. Gschwendtner, S. Mazzoni, A. Petrenko,  CERN, Geneva, Switzerland \\
\\ \textsuperscript{1} also at Technical University of Graz, Graz, Austria}
	
\maketitle

\begin{abstract}
The AWAKE project at CERN uses a self-modulated \SI{400}{GeV/c} proton bunch to drive GV/m wakefields in a \SI{10}{m} long plasma with an electron density of $n_{pe} = 7 \times 10^{14}\,\rm{electrons/cm}^3$. We present the upgrade of a proton beam diagnostic to indirectly prove that the bunch self-modulated by imaging defocused protons with two screens downstream the end of the plasma. The two-screen diagnostic has been installed, commissioned and tested in autumn 2016 and limitations were identified. We plan to install an upgraded diagnostics to overcome these limitations.
\end{abstract}

\section{INTRODUCTION}
The Advanced Proton Driven Plasma Wakefield Acceleration Experiment (AWAKE)\cite{PATHTOAWAKE} is a proof-of-principle R\&D project at CERN. The goal of AWAKE is to accelerate electrons using plasma wakefields with GV/m amplitude created by a self-modulated proton bunch. The proton bunch provided by the CERN SPS has a momentum of \SI{400}{GeV/c}, a longitudinal size $\sigma_z$ of \SI{12}{cm} and a radial beam rms radius of $\sigma_r = $ \SI{0.2}{mm}.  A \SI{100}{fs} long, \SI{450}{mJ} laser pulse ionizes \SI{10}{m} of rubidium vapour and creates a plasma channel with a radius of \SI{1}{mm} and a plasma density ranging from $10^{14}-10^{15}\,\rm{electrons/cm}^3$.

The optimum proton bunch length for a given plasma density is given by $k_{pe} \sigma_z \cong \sqrt{2}$, where $k_{pe} = \omega_{pe}/c$ is the plasma wavenumber, $\omega_{pe}$ the angular plasma frequency and $c$ the speed of light. For a plasma density of $7\times10^{14}\,\rm{electrons/cm}^3$ (AWAKE baseline) the optimum bunch length is in the mm-range, much shorther than the $\sigma_z =$ \SI{12}{cm} of the proton bunch. Hence AWAKE relies on the development of the self-modulation-instability (SMI) \cite{SMI} to modulate the long proton bunch into micro-bunches spaced at the plasma wavelength $\lambda_{pe} = 2\pi/k_{pe}$. The self-modulation instability is seeded by a sharp turn-on of the plasma, realized by placing the ionizing laser pulse in the center (temporally) of the proton bunch.

Seed wakefields periodically focus and defocus protons. As the instability develops, micro-bunches are formed due to the proton density modulation near the axis. These micro-bunches can then resonantly drive the plasma wave. AWAKE studies of the SMI are scheduled for 2016 and 2017, and electron acceleration of externally injected electrons is planned for 2018.

In this article we discuss the upgrade of a proton beam diagnostics to indirectly demonstrate that the self-modulation instability develops within the \SI{10}{m} of plasma.  The basic version of this setup is described in \cite{TURNER 2016} and was successfully operated during a first AWAKE run in December 2016. We explain the limitations we observed with this setup and discuss a major upgrade on the system, to be operational before the next beam time in June 2017.

\section{THE GOAL OF THE INDIRECT SMI TWO-SCREEN MEASUREMENT SYSTEM}
\begin{figure}[htb!]
		\includegraphics[width = 1\columnwidth]{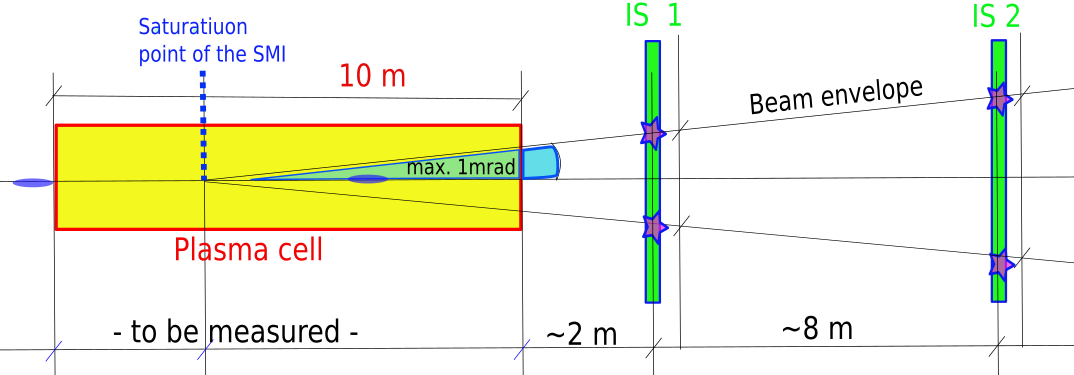}
		\caption{Schematic drawing of the principle behind the two screen measurement setup.}
\label{fig:SMIscetch}
\end{figure}

The basic idea of the two-screen diagnostic is to image the protons that are defocused by the transverse plasma wakefields and to measure their maximum defocusing angle. As indicated in Figure \ref{fig:SMIscetch}, we want to measure maximum radial displacement of the protons by inserting scintillating screens 2 and \SI{10}{m} downstream the plasma. As the protons traverse the screen, the screen emits scintillation light.  The radial, time-integrated proton beam profile is reconstructed by imaging the light onto a camera.

Figure \ref{fig:SMIscetch} also suggests that we can calculate the maximum proton defocusing angle from the two beam images by connecting the measured maximum radial displacements. If SMI develops successfully we expect to measure a maximum defocusing angle around \SI{1}{mrad} instead of the $\approx$ \SI{0.05}{mrad} without plasma and SMI. Defocusing the proton beam from its natural divergence up to $\approx$\SI{1}{mrad} over a distance of $\approx$ \SI{0.5}{m} (as expected from LCODE \cite{LCODE1,LCODE2} simulations) requires GV/m defocusing plasma wakefields.

\begin{figure}[htb!]
		\includegraphics[width = 1\columnwidth]{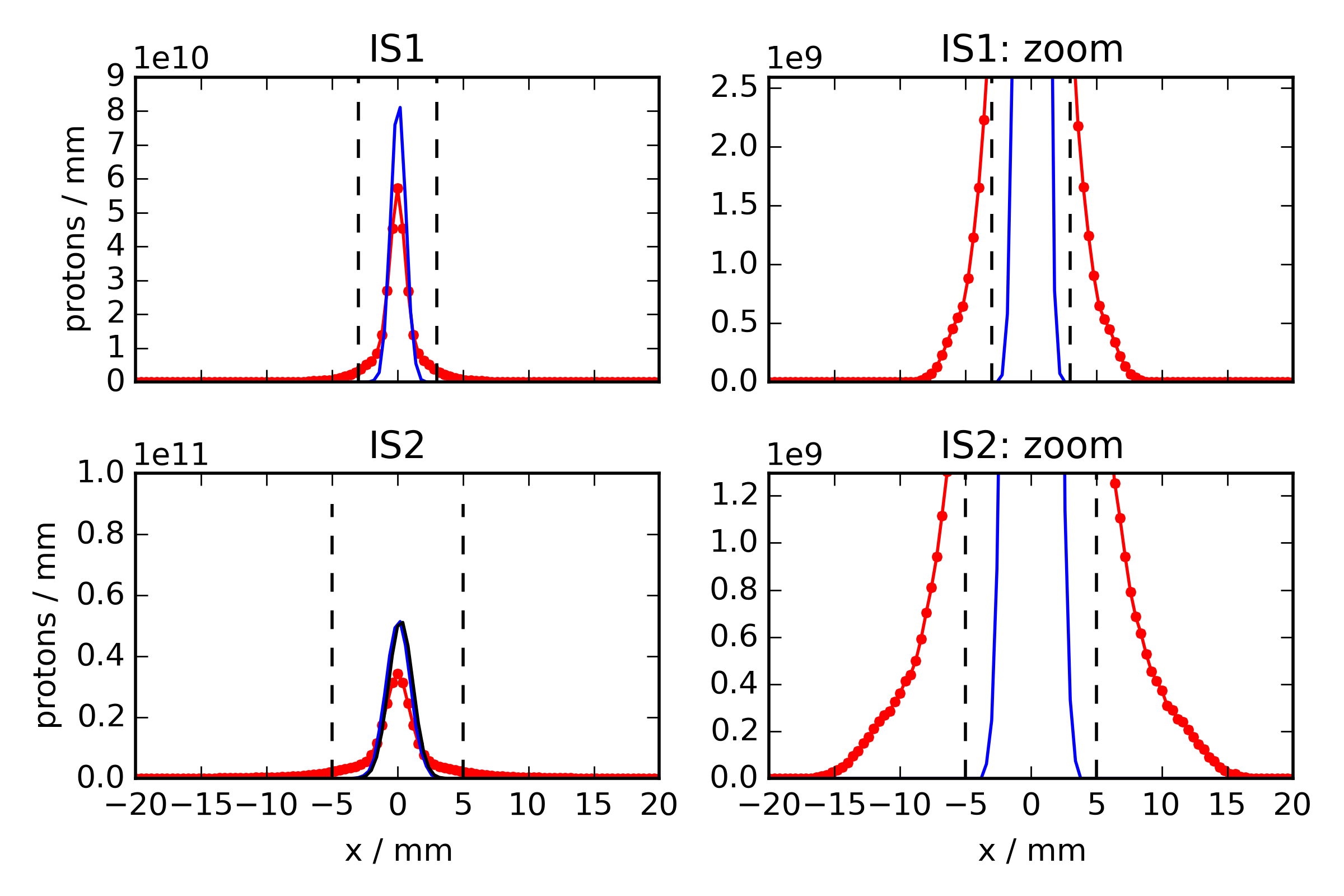}
		\caption{From LCODE \cite{LCODE1,LCODE2} simlations expected transverse profile of the self-modulated proton bunch for the AWAKE baseline parameters \cite{TURNER 2016} (red) and no plasma (blue). Dashed lines indicate the size of the holes that were cut into the imaging screens (see the following section).}
\label{fig:SMIprotonprofile}
\end{figure}

Figure \ref{fig:SMIprotonprofile} shows the expected radial proton beam profiles at the two screen locations in case of no plasma and no SMI (blue line) and with SMI in plasma with a density of $7\times10^{14}\,\rm{electrons/cm}^3$ (red line).

We want to detect the outermost proton beam edges and one major challenge for measuring these edges is that the beam density on the screen ranges over 4-5 orders of magnitude, while the dynamic range of the cameras is limited to 2-3 orders of magnitudes. We previously found that the Chromox screen light yield is directly (linearly) proportional to the proton beam density traversing the screen \cite{IBIC2016}. We do not expect any light yield degradation over time because of the low number of protons per bunch $N_p$ ($N_p = 3\times 10^{11}$) and the low repetition rate (1 bunch every \SI{30}{s}).

\section{THE TWO SCREEN MEASUREMENT SETUP}
Previously, we installed two imaging stations (IS1, IS2) 2 and \SI{10}{m} downstream the end of the plasma (see Figure \ref{fig:SMIscetch}). Each imaging station consisted of a vacuum tank, a scintillating screen (Chromox:Al$_2$O$_3$:CrO$_2$), a filter wheel, a camera lens, a camera and an image readout system.

Since the dynamic range of the camera is limited to 2-3 orders of magnitude, we cut a hole with a radius of 3 and \SI{5}{mm} into the first and the second screen, respectively (see vertical dashed lines in Figure \ref{fig:SMIprotonprofile}). For one screen installed in IS2 we filled the hole with a low light yield material (aluminum) to create a combined screen.
The two imaging stations were commissioned and successfully operated in December 2016, but we identified several limitations of the system:
\begin{itemize}
\item If the experimental observations differ from the expected simulation results shown in Figure \ref{fig:SMIprotonprofile} or we change for example the plasma density, we need to change the size of the hole in the screens. Changing screens is a major intervention that takes several days because it involves breaking the beam-line vacuum.
\item The proton beam has to pass through the center of the hole in the screen with an accuracy of \SI{0.1}{mm}. If the experiment decides to change the proton beam trajectory, the screens have to be realigned. Realignment can be done by precisely moving the vacuum tank, but due to safety reasons can imply breaking the vacuum.
\item The imaging stations use analogue video cameras (standard system at CERN working well in radiation environment). The cameras are triggered every \SI{20}{ms} and integrate over \SI{20}{ms}, independently of the arrival time of the proton bunch. Hence we are unable to compare measurements in terms of signal intensity.
\item Information regarding the beam core is either lost (if the screen has a hole) or limited (combined screen), due to the fixed light yield ratio between the screen materials in the combined screen. 
\end{itemize}
We propose an upgrade to overcome the observed limitations and to make the system more flexible for the experiment.

\section{THE UPGRADED TWO SCREEN MEASUREMENT SYSTEM}
\begin{figure}[htb!]
\centering
		\includegraphics[width =0.8\columnwidth]{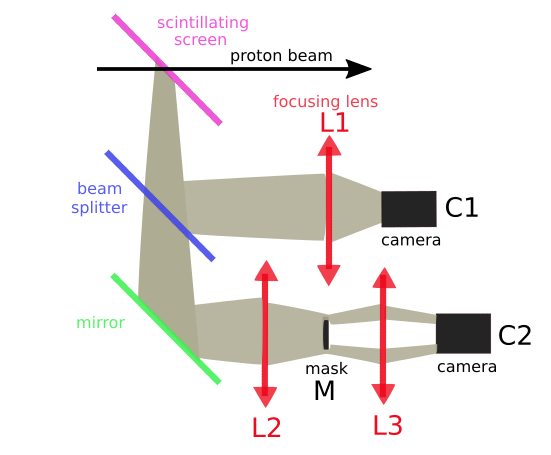}
		\caption{Schematic drawing of the optics setup of the upgraded imaging system of IS1.}
\label{fig:upgradeschematic}
\end{figure}

A schematic of the upgraded optics system for IS1 is shown in Figure \ref{fig:upgradeschematic}.

We replaced the scintillating Chromox screen with a hole by a full Chromox screen. Instead of imaging the light emitted by the scintillator screen directly onto a camera, we split the light using a \SI{3}{inch} pellicle beam splitter that reflects 8\% and transmits 92\% of the emitted light. 

\begin{figure}[htb!]
\centering
		\includegraphics[width =0.8\columnwidth]{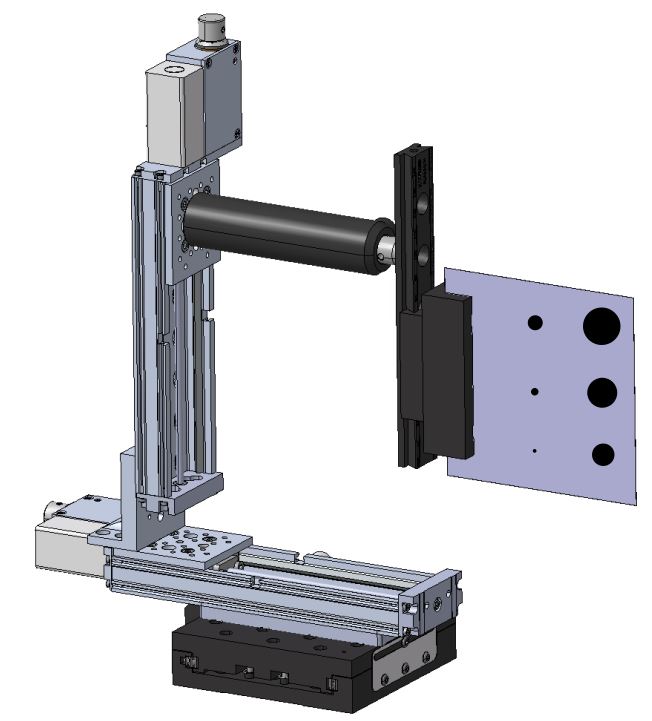}
		\caption{Schematic drawing of the light-core blocking mask system.}
\label{fig:mask}
\end{figure}

The reflected light is directly imaged onto the camera C1 by the lens L1. The image of camera C1 will provide information about the size and position of the beam core. The transmitted light is imaged by lens L2 (\SI{3}{inch}, f=\SI{200}{mm}) onto the plane M, where the mask is placed. The mask blocks the intense beam core by an opaque region in mask. A technical drawing of the mask setup is shown in Figure \ref{fig:mask}. The mask is reimaged by lens L3 onto camera C2.

The mask system consists of a \SI{1}{mm} thick fused silica window with different sized opaque coatings, a glass holder, two stepper motors for horizontal and vertical alignment and a manual translation stage to faciliate imaging by longitudinal positioning on the table. The distances are chosen to have an image magnification of 0.5 in the mask plane. The light passing around the mask is then reimaged onto an externally triggered \SI{12}{bit} camera (Basler ac1600-60gm).

The optics for IS2 are very similar to the one from IS1 (see Figure \ref{fig:upgradeschematic}), but due to space restrictions the beam splitter comes after the mirror.

We have several different mask sizes available and we can remotely adjust and align  the mask for various cases (i.e. plasma densities, changing proton beam trajectory, etc.) by varying the radius to which we want to block the beam core.
We use a camera, that is synchronized to the proton beam -to capture all the emitted light- so that we can directly compare different events and explore the radial proton beam profile in details. We can now optimally observe the beam core and the defocused protons in terms of light intensity and mask size.

The procedure will be as follows: We locate the beam without mask, align the mask onto the beam and then adjust filters in the optical line to get the right exposure of the camera.

We previously found \cite{TURNER 2016} that the main background contribution comes from secondary particles directly impacting onto the camera. The laser dump $\approx$ \SI{30}{cm} upstream IS1 shields the imaging screen from low-energy electrons from the plasma. Since the plasma wakefield amplitude is low in this experiment, trapping and accelerating electrons is unlikely.

\section{SUMMARY}
We discussed an upgrade of the indirect SMI two-screen measurement system to be installed in the AWAKE experiment and ready for operation in June 2017. The idea of the two screen measurement setup is to image protons that got defocused by the transverse plasma wakefields to indirectly show that the proton bunch self-modulated and wakefields were created in the \SI{10}{m} of plasma. The upgraded system offers different mask sizes to block the light from the beam core, facilitates alignment of the masks with respect to the proton beam, enables comparability between measurements and images also the proton beam core.

In the upgraded system we split the light emitted by the scintillator screen and image one part directly onto a camera to obtain information about the beam core. The other part blocks the beam core with a mask with several mask sizes available, to detect the maximum defocused protons.

\null
\end{document}